\documentclass[useAMS,usenatbib]{mn2e}
\usepackage{times,graphicx,colordvi}
\newcommand{\apj}{ApJ}                 
\newcommand{\mnras}{MNRAS}             \newcommand{\aap}{A\&A}
             
\newcommand{\aj}{AJ}

\newcommand{\dd}{{\rm d}}
\newcommand{\vc}[1]{\textbf{\emph #1}}
\begin{document}
\title[Statistical mechanics of self-gravitating systems]{Fluid-like entropy and equilibrium statistical mechanics of self-gravitating systems}
\author[D.-B. Kang and P. He]{Dong-Biao Kang\thanks{E-mail: billkang@itp.ac.cn} and Ping He \\
Key Laboratory of Frontiers in Theoretical Physics, Institute of Theoretical Physics, Chinese Academy of Science, Beijing 100190, China}
\date{\today}
\maketitle
% ------------------------------------------------------------------------------------
\begin{abstract}
The statistical mechanics of self-gravitating systems has not been well understood, and still remains an open question so far. In a previous study by Kang \& He, we showed that the fluid approximation may give a clue to further investigate this problem. In fact, there are indeed many dynamical similarities between self-gravitating and fluid systems. Based on a fluid-like entropy, that work explained successfully the outer density profiles of dark matter halos, but there left some drawbacks with the calculation concerning extremizing process of the entropy. In the current paper, with the improved extremizing calculation -- including an additional differential constraint of dynamical equilibrium and without any other assumptions, we confirm that statistical-mechanical methods can give a density profile with finite mass and finite energy. Moreover, this density profile is also consistent with the observational surface brightness of the elliptical galaxy NGC 3379. In our methods, the density profile is derived from the equation of state, which is obtained from entropy principle but does not correspond to the maximum entropy of the system. Finally, we suggest an alternative entropy form, a hybrid of Boltzmann-Gibbs and Tsallis entropy, whose global maximum may give rise to this equation of state.
\end{abstract}

\begin{keywords}
methods: analytical -- galaxies: structure -- equation of state -- galaxies: kinematics and dynamics.
\end{keywords}

% --------------------------------------------------------------------------------------
\section{Introduction}
\label{s1}

We know that two-body relaxation may play a crucial role in systems such as stellar clusters and the center of galaxies, while plays no role in collisionless systems such as galaxies and dark mater halos, whose time-scales of two-body relaxation are longer than the age of the universe. \citet{lb67} proposed `violent relaxation', which is another relaxation mechanism due to the rapidly changing gravitational field when systems collapse. Violent relaxation can allow the collisionless system to be in equilibrium. If we assume that the Boltzmann-Gibbs (BG) entropy is applicable to self-gravitating systems
% --------------------------------------------------------------------------------------
\begin{equation}
\label{BG}
S = - \int f \ln f \dd \tau,
\end{equation}
where $f=f(\vc{x}, \vc{v})$ is the distribution function, and $\dd\tau = \dd^3 \vc{x} \dd^3 \vc{v}$ is the phase-space element. By maximizing the BG entropy with the constraints of given mass and energy, we can obtain the Maxwellian distribution for the system, which corresponds to an isothermal density profile and hence cannot be accepted because of its infinite mass and infinite energy. Then, to eliminate this inconsistency, \citet{lb67} also proposed the incomplete relaxation, which means that the isothermal solution is only valid for the central region.

In fact, the reason for this inconsistency is a result pointed out by \citet{ant62} that the maximum of the BG entropy of self-gravitating systems does not exist, so the thermodynamic equilibrium state may not exist. Binney \citep[see][]{galdyn08} also proved that if one can divide the system into a central main body and an outer envelope, then the BG entropy of self-gravitating systems can increase without upper bound by increasing the system's central concentration. So there is no global maximum of entropy for self-gravitating systems, or we can question the validity of the BG entropy for self-gravitating systems.

Despite these difficulties, statistical-mechanical methods have always been explored in an attempt to explain the properties of self-gravitating systems. \citet{tremaine86} discussed the generalized $H$-function
% -------------------------------------------------------------------------------------
\begin{equation}
\label{convex}
H[f] = -\int C(f)\dd\tau,
\end{equation}
where $C(f)$ is any convex functional of $f$. In fact, the BG entropy (\ref{BG}) is just one of such $C(f)$s and may be only valid for short-range interacting systems. For long-range interacting systems, such as self-gravitating systems, \citet{tsallis88} generalized the form of BG entropy in mathematics as
% --------------------------------------------------------------------------------------
\begin{equation}
S_q[f] = -\frac{1}{q-1}\int(f^q-f) \dd\tau,
\end{equation}
and established the non-extensive statistical mechanics. When $q=1$, it recovers to BG entropy. \citet{han04} used this entropy to investigate the density slope of dark matter halos. However, the maximum of this entropy cannot give a correct density profile of self-gravitating system. Some authors also studied the thermodynamics and kinetic equations of long-range systems, such as \citet{chavanis06} and \citet{chavanis10}.

Besides, some works also applied some special entropy forms to self-gravitating systems. \citet{white87} (hereafter, WN87) considered the specific entropy form of ideal gas. With fixed mass, fixed energy and finite extent, WN87 seeks the equilibrium configuration of elliptical galaxies by maximizing the entropy of ideal gas, but the result seems to be not good. However, in this paper we will argue about the validity of their calculation. Our previous works \citet{hep10} and \citet{kang11} (hereafter, KH11) employed this entropy form to study the equilibrium configurations of dark matter halos. Moreover, KH11 considered the general anisotropic situation and found an equation of state whose density profile has finite mass and extent. Here we will also discuss main properties of this density profile.

In this paper, we demonstrate that the results of WN87 and KH11 should have no essential differences and propose another kind of entropy form including both BG entropy and Tsallis Entropy. This paper is organized as follows. In section~\ref{s2}, we emphasize the dynamical similarities between self-gravitating systems and fluids. We restate the work of KH11 with some differences and mainly compare the results of WN87 and KH11. In section~\ref{s3} we propose the new entropy form of the hybrid BG and Tsallis entropy. We present the conclusion of our work in section~\ref{s4}. The appendix \ref{a1} explains the derivation of the new equation of state which appears in section~\ref{s2.2}.

% --------------------------------------------------------------------------------------
\section{Basic theory and formulae}
\label{s2}

% --------------------------------------------------------------------------------------
\subsection{Connections between fluids and self-gravitating systems}
\label{s2.1}

For collisional systems, some authors \citep[e.g.,][]{lar70,lyn80,bett83} integrate the Boltzmann equation to obtain a set of moment equations, which can be solved to follow the evolution of the system. Under some reasonable assumptions, only the low-order moment equations are reserved and become fluid-like conservation equations. So the collisional gravitating systems may be approximated by fluids whose hydrodynamic equilibrium can be shown by
% --------------------------------------------------------------------------------------
\begin{equation}
\label{hdr}
\nabla p = -\rho\nabla\Phi,
\end{equation}
where $\Phi$ and $p$ are the gravitational potential and pressure respectively, and the density $\rho$ satisfies the Poisson equation
% -------------------------------------------------------------------------------------
\begin{equation}
\label{phi}
\nabla^2\Phi = 4\pi G\rho.
\end{equation}
While the collisionless system is described by the phase-space density $f=f(\vc{x}, \vc{v})$ which satisfies the collisionless Boltzmann equation \citep[CBE, see][]{galdyn08}. For simplicity, we restrict ourselves to the spherical systems in this work. Then \citet{ahn05} \citep[also see][]{Teyssier97,Subramanian00} proved that with the three assumptions of spherical symmetry, skew-free velocity distribution and isotropic velocity dispersion, we can also integrate CBE as the case for collisional systems, and the derived moment equations are the same as fluid conservation equations for a gas with adiabatic index $5/3$. When the collisionless systems are in dynamical equilibrium, we can get the Jeans equation from CBE:
% --------------------------------------------------------------------------------------
\begin{equation}
\label{jeans}
\frac{\dd(\rho\sigma_r^2)}{\dd r} + 2\beta\frac{(\rho\sigma_r^2)} {r} = -\rho \frac{\dd\Phi} {\dd r},
\end{equation}
where $\rho$ satisfies equation~(\ref{phi}), with
% -------------------------------------------------------------------------------------
\begin{equation}
\label{rho}
\rho(\vc{x})=\int f(\vc{x}, \vc{v}) \dd^3 \vc{v},
\end{equation}
and $\beta(r) = 1 - \sigma_t^2/2\sigma_r^2$ being the anisotropy parameter, and $\sigma_t^2$ and $\sigma_r^2$ are the tangential and radial velocity dispersion, respectively. As in KH11, we define the effective pressure as
% --------------------------------------------------------------------------------------
\begin{equation}
\label{pdef}
P = -\int^{\infty}_r \big(\frac{\dd(\rho\sigma_r^2)} {\dd r'} + 2\beta \frac{(\rho\sigma_r^2)}{r'}\big)\dd r'.
\end{equation}
Then equation~(\ref{jeans}) can be rewritten as
\begin{equation}
\label{eff}
\frac{\dd P}{\dd r}=-\rho\frac{\dd\Phi}{\dd r}=-\rho\frac{G M}{r^2},
\end{equation}
which is formally equivalent to equation~(\ref{hdr}). Hence, with the definition of the effective pressure $P$, such a formal equivalence suggests that there should be some dynamical similarities between self-gravitating systems and fluids.

% --------------------------------------------------------------------------------------
\subsection{Equation of state}
\label{s2.2}

We will reproduce the equation of state in KH11 with the improved methods of the variational calculus. Based on the analysis of the above section, we infer that the fluid-like entropy with the adiabatic index $5/3$ is really applicable to current self-gravitating systems as
% ------------------------------------------------------------------------------------
\begin{equation}
\label{entropy}
S_t = \int_0^{\infty} \rho s 4\pi r^2 \dd r = \int_0^{\infty} \rho\ln(\frac {P^{3/2}} {\rho^{5/2}}) 4\pi r^2 \dd r,
\end{equation}
where $s=\ln(P^{3/2}/\rho^{5/2})$ is defined in KH11. Setting $y(r)=M(r)=\int_0^r4\pi\rho r^2 \dd r$, the mass function of the system, we obtain
% -------------------------------------------------------------------------------------
\begin{equation}
\label{density}
\rho=\frac{y'}{4\pi r^2}.
\end{equation}
The mass constraint for extremizing the entropy is equivalent to a fixed end-point condition as
% -------------------------------------------------------------------------------------
\begin{equation}
M_t=y(r)|_{r=\infty}.
\end{equation}
The energy constraint is
% -------------------------------------------------------------------------------------
\begin{equation}
\label{energy}
\begin{array}{ll}
E_t = \displaystyle E_k + E_v = \frac{1}{2}\int\rho\sigma_{tot}^2 \dd V -\int \frac{G\rho(\vc{r}) \rho (\vc{r}')}{2|\vc{r}-\vc{r}'|}\dd V \dd V' \\
= \displaystyle 4\pi\int_0^{\infty} \big(\frac{(3-2\beta)}{2}\rho\sigma_r^2r^2 - G M \rho  r \big)\dd r \\
=\displaystyle\int_0^{\infty}\big(6\pi r^2P-\frac{G y y'}{r}\big)\dd r,
\end{array}
\end{equation}
in which we used a change of integration order to derive the first term in the last line of the above equation. It is different from KH11, in that (1) we have not assumed that the system is in hydrostatic equilibrium, so the virial theorem cannot be used; (2) $P$ is treated as an independent variable rather than a known function. Then the constrained entropy will be
% --------------------------------------------------------------------------------------
\begin{equation}
\label{variation}
S_{t,r} = S_t + \lambda E_t,
\end{equation}
where $\lambda$ is a Lagrangian multiplier. Substituting equations~(\ref{entropy}), (\ref{density}) and (\ref{energy}) into (\ref{variation}), we obtain
% -------------------------------------------------------------------------------------
\begin{equation}
\begin{array}{ll}
S_{t,r} = \displaystyle\int_0^{\infty} F(r,y,y',P,P')\dd r\\
\displaystyle = \int_0^{\infty} \big( y'\ln(P^{3/2}(\frac{y'}{4\pi r^2})^{-5/2} +\displaystyle\lambda(6\pi r^2P-\frac{Gyy'}{r})\big) \dd r.
\end{array}
\end{equation}
We use entropy principle, i.e. $\delta S_{t,r}=0$, to extremize $S_{t,r}$ and obtain the Euler-Lagrangian equation for $P$ and $y$ respectively, as
% --------------------------------------------------------------------------------------
\begin{equation}
\label{var1}
\rho=-\lambda P,
\end{equation}
and
% --------------------------------------------------------------------------------------
\begin{equation}
\label{var2}
\frac{\dd\ln(P^{\frac{3}{2}}/\rho^{\frac{5}{2}})}{\dd r} = -\lambda\frac{G M} {r^2}.
\end{equation}
In combination with equation~(\ref{eff}), the solution of equations~(\ref{var1}) and (\ref{var2}) is an isothermal sphere, which does not require that $\beta=0$. However, this solution cannot be accepted, because (1) as mentioned previously, it has infinite mass and energy; (2) it has only one shape parameter $\lambda$ but there are two constraints, and thus this solution is incomplete. Nevertheless, if we discard equation~(\ref{var1}) and only substitute equation~(\ref{eff}) into the RHS of equation~(\ref{var2}), we can get the result of KH11
% --------------------------------------------------------------------------------------
\begin{equation}
\label{eqsta}
\rho = -\lambda P + \nu P^{3/5},
\end{equation}
where $\nu$ is an integral constant.

\begin{figure}
\centerline{\includegraphics[width=\columnwidth]{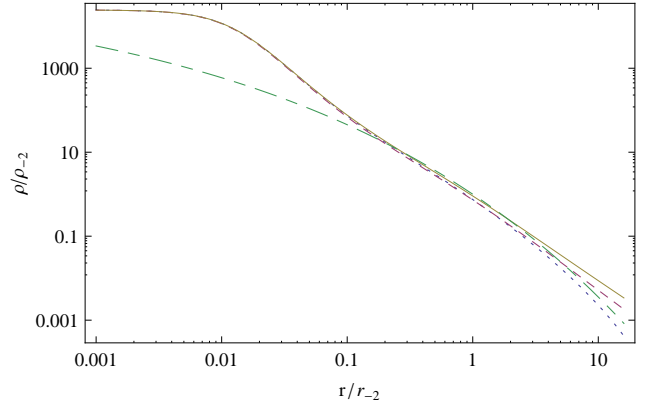}}
\caption{Density profiles of different equations of state. The solid, dashed and dotted lines correspond to equations~(\ref{var1}), (\ref{sta:gal}) and (\ref{eqsta}), respectively. The long-short dashed line indicates the Einasto profile with $\alpha=0.17$. Here we take $\lambda=-2.27$, $\nu=0.5$ and set $4\pi G = \rho_{-2}=r_{-2}=1$, where $r_{-2}$ is the radius when $\dd\ln\rho/\dd\ln r=-2$.}
\label{dens1}
\end{figure}

Here we summarize the main properties of equation~(\ref{eqsta}) shown by KH11:
\begin{enumerate}
\item There are two shape parameters $\lambda$ and $\nu$ that can be adjusted to make equation~(\ref{eqsta}) satisfy the mass and energy constraints, and the existence of $\nu$ term amounts to introducing a `truncated' radius, which ensures the finite mass of the gravitating system.
\item $\lambda$ must be negative and the $\nu$ term may play an important role only at large radii. The density profile is shown in Figure~\ref{dens1}, which is similar to a `truncated' cored isothermal sphere. We can get different core radii and the `truncated' radii by specifying different values of $\lambda$ and $\nu$. As shown in Figure~\ref{dens1}, this density profile is only consistent with the dark matter halo's Einasto profile \citep{navarro10} in the range of $r/r_{-2}>0.2$. Nonetheless, this solution should be acceptable.
\end{enumerate}

According to these results, we can identify an interesting structure, i.e. index-varying polytropic system, with the polytropic index (see equation~(\ref{pol}) and below) changing from $1$ to $5/3$ with $r$ increasing from $0$ to $\infty$, and correspondingly, the state of the system varies from isothermal to adiabatic, which is different from the `core-halo' structure. However, this structure may not correspond to the extremum of the entropy, which may be a severe problem. A possible explanation to this problem is as follows. The relaxation may be incomplete \citep{lb67}, and only the central part of system is sufficiently relaxed. Hence the central part is at the extremum of the entropy and isothermal, while the outmost part in fact can be treated as being isolated from the whole system, and thus it is adiabatic.

Then, if we treat equation~(\ref{eff}) as a differential constraint to the above variational calculus, similar to the manipulations of WN87, the constrained entropy becomes
% --------------------------------------------------------------------------------------
\begin{equation}
\label{entropy2}
\begin{array}{lll}
S_{t,r} = \displaystyle\int_0^{\infty} F(r,\eta,y,y',P,P')\dd r \\
= \displaystyle\int_0^{\infty} \left( y'\ln\big(P^{3/2}(\frac{y'}{4\pi r^2})^{-5/2}\big) - \lambda \frac{G yy'}{2r} \right.\\
+ \displaystyle \left. \eta(r)(\frac{\dd P}{\dd r}+\frac {G y y'}{4\pi r^4}) \right) \dd r,
\end{array}
\end{equation}
where virial theorem has been used because now we have constrained the system to satisfy equation~(\ref{eff}), and $\eta$ is a Lagrangian multiplier. Performing a similar variational calculus, we can reproduce the similar result as WN87,
% --------------------------------------------------------------------------------------
\begin{equation}
\label{eqsta2}
\eta'=\frac{6\pi r^2\rho}{P}, {\hskip 5mm} \frac{\dd\rho} {\dd r} = -\frac{\dd P} {5\dd r} (\lambda - \frac{2 \eta}{\pi r^3}).
\end{equation}
The problem is that equation~(\ref{eqsta}) and (\ref{eqsta2}) are both obtained by using entropy principle with the constraint of hydrodynamical equilibrium, and thus we speculate that the equation of state (\ref{eqsta2}) should derive similar density profile as equation~(\ref{eqsta}). However, KH11 produces an index-changing polytropic structure, while WN87 exhibits a core-halo structure. We provide an explanation to this apparent contradiction in Appendix~\ref{a1}, where we also derive an approximated solution of equation~(\ref{eqsta2})
% --------------------------------------------------------------------------------------
\begin{equation}
\label{sta:gal}
\rho = -\lambda P + \nu P^{4/5}.
\end{equation}
Then equations of state (\ref{eqsta}) and (\ref{sta:gal}) are found to have similar forms, and their solutions can give us similar density profiles which are shown in Figure~\ref{dens1}. So it is not necessary to introduce the incomplete relaxation here.

% -------------------------------------------------------------------------------------

% -------------------------------------------------------------------------------------
An advantage of equation of state (\ref{eqsta2}) is that it is at the extremum of the entropy, with which we can further study self-gravitating systems using statistical-mechanical methods. Then we calculate the second order variation of entropy in equation~(\ref{entropy2}) as
% --------------------------------------------------------------------------------------
\begin{equation}
\label{2dervar}
\begin{array}{ll}
\delta^2S_{t,r} = \displaystyle\frac{1}{2} \int\big((-\lambda \frac{G}{2r^2} - G (\frac{\eta}{r^4})')(\delta M)^2 \\
\displaystyle-\frac{10\pi r^2}{\rho}(\delta\rho)^2 -\frac{6 \pi r^2\rho} {P^2} (\delta P)^2\big) \dd r,
\end{array}
\end{equation}
where the coefficient of $(\delta M)^2$ can be positive because $\lambda<0$. So according to the calculus of variations, our equation~(\ref{sta:gal}) does not correspond to the global maximum of the constrained entropy of equation~(\ref{entropy2}). This result is consistent with one of our earlier works, in which we find that the equilibrium states of self-gravitating systems are the saddle-point entropy states \citep{hep11}.

% --------------------------------------------------------------------------------------
\subsection{Comparison with observations}
\label{s2.3}

\begin{figure}
\centerline{\includegraphics[width=\columnwidth]{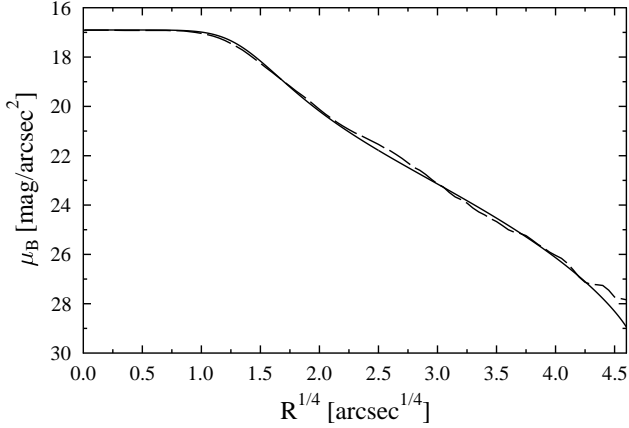}}
\caption{The surface brightness profile (projected density profile) of the elliptical galaxy. The dashed curve comes from the measurement of NGC 3379 \citep{capa90}, and the solid line is calculated from equation~(\ref{sta:gal}) for a specific choice of the values of $\lambda$ and $\nu$. See \citet*{mo10} for details about the calculation of $\mu_B$.}
\label{dens2}
\end{figure}

Coincidentally, equation~(\ref{sta:gal}) is very similar to the result of a model proposed by \citet{jaffe87} and \citet{hm93}, who assumed that
% --------------------------------------------------------------------------------------
\begin{equation}
N(\epsilon )\approx \left\{
\begin{array}{ll}
0{\hskip 3mm}\texttt{for }\epsilon \geqslant 0\\
N_0{\hskip 2mm}\texttt{for }\epsilon \rightarrow  0^-
\end{array}
\right.,{\hskip 3mm}f(\epsilon )\sim e^{-\beta \epsilon },
\end{equation}
where $N(\epsilon)$ is the differential energy distribution, then the density profile is described by an isothermal core and a polytropic halo with index $5/4$. According to the result of \citet{jaffe87}, this model in fact uses an approximation that $\Phi\sim-GM/r$ at large radii, while equation~(\ref{eqsta}) means that the outmost part is isolated from the system, i.e. $\Phi\sim0$ at large radii, so equation~(\ref{sta:gal}) is more accurate than equation~(\ref{eqsta}). According to the result of \citet{hm93}, their model can fit the surface photometry along the major axis of NGC 3379 very well, and we also present one solution of equation~(\ref{eqsta2}) in Figure~\ref{dens2}, which is compared with the density profile of NGC 3379. Besides, as addressed previously, equation~(\ref{sta:gal}) can be approximated as the truncated isothermal sphere, which is just the King model \citep{King66}, then our results are also applicable to globular clusters. Hence it is interesting to see that equation~(\ref{sta:gal}) accounts for observations much better than for the simulation results concerning dark matter halos, especially in the center part of the system.

% --------------------------------------------------------------------------------------

% -----------------------------------------------------------------------------------
\section{Form of the entropy}
\label{s3}

In the above section, although we get a density profile with finite mass using statistical-mechanical methods, the extremum of entropy is not a global maximum, which is not consistent with the usual principle of maximum entropy. We consider that it may be still necessary to obtain equation~(\ref{eqsta2}) by maximizing some specific entropy form, which is constructed from the distribution function $f$.

From observations and numerical simulations \citep[e.g.][]{navarro10,guimaraes11}, we know that the velocity distributions of stellar systems and dark matter halos are anisotropic, while BG entropy maximization with mass and energy constraints gives isotropic distribution function. Some studies \citep{bertin03,bertin05} considered the derivation of anisotropic distribution functions from entropy principle. They calculate the extremum of the BG entropy by imposing the constraints of mass, energy and a third quantity $Q$, which is related to the single particle's angular momentum and is introduced to consider the incomplete violent relaxation. But here we do not consider the anisotropy and only impose energy and mass constraints on the entropy maximization, so we will derive a density profile from an isotropic distribution function. Then notice that through the definition of the effective pressure we can always get a relation among $\beta$, $\sigma_r^2$ and $\rho$ (see KH11), but this relation will not be discussed again here.

It is interesting to notice that the extremum of BG entropy leads to an isothermal solution, while the extremum of Tsallis entropy gives rise to a polytropic solution \citep[e.g.][]{han04}, and equation~(\ref{sta:gal}) is just a sum of isothermal and polytropic solutions; while BG entropy describes short-range interacting systems and Tsallis entropy may describe long-range interacting systems. These suggest that the proper entropy for self-gravitating systems, which is a functional of $f$, may be a hybrid of BG entropy and Tsallis entropy. Notice that although collisionless gravitating systems are described by CBE, we can never neglect collisions at least in the central regions ($r\sim0$) of the real systems. For these reasons, we consider an entropy form as
% --------------------------------------------------------------------------------------
\begin{equation}
\label{entropy3}
S[f]=S[f_1,f_2]=\int (-f_1\ln f_1-\frac{f_2^q-f_2}{q-1})\dd\tau,
\end{equation}
where $f_1 = \omega f, f_2 = (1 - \omega)f$, $0 < \omega < 1$ and $\omega$ is a parameter. Notice that equation~(\ref{entropy3}) is consistent with equation~(\ref{convex}). The constraints are:
% --------------------------------------------------------------------------------------
\begin{equation}
\int f \dd\tau=M, {\hskip 5mm} \int f\epsilon \dd\tau=E,
\end{equation}
where $\epsilon = v^2/2+\Phi$. The constraint of hydrodynamics equilibrium can be realized by $f=f(\epsilon)$ \citep{chavanis08}. Then the constrained entropy $S_t$ will be:
% --------------------------------------------------------------------------------------
\begin{equation}
\label{var3}
S_t=\int [-f_1\ln f_1-\frac{f_2^q-f_2}{q-1} + \alpha(f_1+f_2) + \lambda(f_1+f_2) \epsilon] \dd\tau,
\end{equation}
where $\alpha$ is also a Lagrangian multiplier. Evidently, from $\delta S_t=0$ we get:
% --------------------------------------------------------------------------------------
\begin{equation}
\label{eqps}
\omega\ln(\omega f) + (1 - \omega) \frac{q(f - \omega f)^{q-1}}{q-1}-\lambda\epsilon'=0,
\end{equation}
where $\epsilon'=v^2/2+\Phi+[(q-\omega)\alpha-\omega]/\lambda(q-1) = v^2/2 + \Phi'$, and we will leave out the prime in the following text. The stability of the system requires that $f'(\epsilon)<0$ \citep{galdyn08}, so $\lambda<0$. If $\omega = 1$, we can get the result of the BG entropy, so $\rho = -\lambda P$; if $\omega$ approaches $0$, from equation~(\ref{rho}) we have
% --------------------------------------------------------------------------------------
\begin{equation}
\label{pol}
\begin{array}{ll}
\displaystyle\rho=4\pi(\frac{(1-q)\lambda}{q})^{\frac{1}{q-1}} \int_0^{\sqrt{-2\Phi}} v^2 (-\epsilon)^{\frac{1}{q-1}} \dd v \\
\displaystyle = c_n(-\Phi)^{n} = \frac{1}{\kappa}P^{\frac{1}{\gamma}},
\end{array}
\end{equation}
where
% --------------------------------------------------------------------------------------
\begin{displaymath}
c_n=2^{n+2}\pi(\frac{(1-q)\lambda}{q})^{\frac{1}{q-1}} \int_0^{\frac{\pi}{2}}{\sin}^2 \theta{\cos}^{\frac{q+1}{q-1}} \theta \dd\theta,
\end{displaymath}
\begin{equation}
\label{coefficient}
\gamma=1+\frac{1}{n}, n=\frac{3q-1}{2(q-1)}, \kappa=\frac{\gamma-1}{\gamma}c_n^{1-\gamma},
\end{equation}
in which $\gamma$ is the polytropic index. The upper limit of the integral in equation~(\ref{pol}) is $\sqrt{-2\Phi}$, because we will ensure that $\epsilon<0$ for general $q$. In fact, only we take this upper limit, can we get the last step of equation~(\ref{pol}), so we think that this upper limit is responsible for finiteness of the mass of the self-gravitating system. Against our expectation, we cannot get equation~(\ref{sta:gal}) from equation~(\ref{eqps}) when $q=7/5$ (and hence $\gamma = 5/4$).

Then we consider another case, that is, the entropy form is the same as equation~(\ref{entropy3}), but we assume that $f_1, f_2$ will be treated as independent variables so that $\delta S_t=0$ should hold for $f_1$ and $f_2$, respectively. With the following two assumptions that $P_1=P_2=P$ and $\rho=\omega\rho_1 + (1-\omega)\rho_2$ ($P$ is determined by $f$ through equations~(\ref{rho}) and (\ref{pdef})), we derive the result of the extremum of the entropy:
% --------------------------------------------------------------------------------------
\begin{displaymath}
\rho_1=-\lambda P_1, {\hskip 5mm} \rho_2=\frac{1}{\kappa}P_2^{\frac{1}{\gamma}},
\end{displaymath}
so $\rho=-\omega\lambda P+(1-\omega)P^{1/\gamma}/\kappa=-\lambda'P+\omega'P^{1/\gamma}$, which must be equivalent to equation~(\ref{sta:gal}) when $q=7/5$. We make these assumptions mainly due to the following two reasons: (1) the condition $P_1 = P_2 = P$ can ensure $E_{k1}=E_{k2}=E_k$ through equation~(\ref{energy}), and thus $f_1$, $f_2$ and $f$ all satisfy the constraint of the energy, i.e. $E_1=E_2=E$; and (2) $\rho=\omega\rho_1+(1-\omega)\rho_2$ indicates that the spatial density may be a combination of the two terms corresponding to $f_1$ and $f_2$, respectively.

For this case the second order variation of the constrained entropy of equation~(\ref{var3}) is
% --------------------------------------------------------------------------------------
\begin{equation}
{\delta}^2 S_t=\frac{1}{2}\int(-\frac{1}{f_1}(\delta f_1)^2-q{f_2}^{q-2}(\delta f_2)^2) \dd\tau,
\end{equation}
which implies that $f_1$ increases the BG entropy, $f_2$ increases the Tsallis entropy, and hence $S$ is globally maximized.

% --------------------------------------------------------------------------------------
\section{Conclusion}
\label{s4}

The final statistical equilibrium states of isolated systems can be obtained by calculating the extremum of the entropy with the constraints of fixed mass and energy. Based on the dynamical similarities between self-gravitating systems and fluids, we further confirm the validity of fluid-like entropy for self-gravitating systems. Also, we improve the variational calculation of our previous work \citep{kang11}, and propose an index-varying polytropic structure, which has finite mass and can be obtained under the assumption of incomplete relaxation. If we treat the equation of dynamical equilibrium as a differential constraint to the entropy, we derive the similar equation as \citet{white87}. Hence, we conclude that with statistical-mechanical methods we can obtain a density profile with finite mass without introducing the incomplete relaxation. It is interesting to notice that our results may be more consistent with observations than numerical simulations. Insisting on the validity of the principle of maximum entropy, and based on the new equation of state derived in this paper, we suggest a specific entropy form, which is just a hybrid of Boltzmann-Gibbs entropy and Tsallis entropy. We argue that this hybrid entropy may completely describe self-gravitating systems with both short-range and long-range interactions.

% --------------------------------------------------------------------------------------
\section*{Acknowledgements}
DBK is grateful for many helpful suggestions and comments by the anonymous referee. This work is supported by the National Basic Research Programm of China, NO:2010CB832805.

% --------------------------------------------------------------------------------------

% --------------------------------------------------------------------------------------
\appendix\section{}
\label{a1}

Here we present our understandings of the solution of equation~(\ref{eqsta2}), which is different from WN87. WN87 in their figure~2 showed the numerical solution of equation~(\ref{eqsta2}), from which we speculate that their solution has no essential difference from the cored isothermal sphere ($\rho=-\lambda P$), which is an incomplete solution and has infinite mass. But we consider that isothermal sphere is not the only solution of equation~(\ref{eqsta2}) because of two mathematical reasons:
% -------------------------------------------------------------------------------------
\begin{enumerate}
\item We set $m(r)=3r^{-3}\int_0^r r^2\rho/ P dr = \eta(r)/2\pi r^3$. If $m(r)\equiv const$, we must have $\rho/P\equiv const$. Then from equation~(\ref{eqsta2}) we know $\rho=-\lambda P$ and $m=-\lambda$. However, there is no physical requirement that $m(r)\equiv const$, i.e. $m(r)$ is commonly not a constant, so more general solutions must exist.
\item We can rewrite equation~(\ref{eqsta2}) as one differential equation by eliminating $\eta(r)$
% ----------------------------------------------------------------------------------
\begin{equation}
\label{second-order}
5\frac{\dd^2\rho}{\dd P^2}P'(r)r + 15\frac{\dd\rho}{\dd P} + 3\lambda = 12 \frac{\rho}{P},
\end{equation}
where $P'(r)$ is the derivative of $P$ with respect to $r$. Because $P=P(\rho,r)$, we can treat $r$ and $P'(r)$ as functions of $\rho$ and $P$. Then equation~(\ref{second-order}) is a second-order differential equation of $\rho$ with respect to $P$, so there must be two integral constants in the solution $\rho=\rho(P)$. The only physically reasonable boundary condition that we used in this paper is $P=0$ when $\rho=0$, which was already used by WN87, and thus there is only one integral constant left. However, WN87 also set $P|_{r=0}=\rho|_{r=0}=1$, which eliminates the only remaining integral constant and is not proved to be reasonable. This is the reason why WN87 ``comes to an important and surprising point" and ``contains no model with a concentration larger than 3.35". WN87 considered that another shape parameter should be the minimal radius they introduced, but this quantity may never be treated as an integral constant. Hence the solution $\rho=\rho(P)$ must have two shape parameters, including a Lagrangian multiplier $\lambda$ and an unknown integral constant.
\end{enumerate}
Because $m(r)$ is an arbitrary function of $r$, the general exact solutions are very difficult to obtain, and thus we consider to use the iterative method to look for the solution. We first assume that $m(r)\approx\rho(r)/P(r)$, since (1) when $r\rightarrow0$ and $r\rightarrow\infty$, then $m(r)\rightarrow\rho/P$ according to the L'H\^{o}pital's rule; and (2) $m(r)=\rho(r)/P(r)$ can exactly set up only when $m\equiv- \lambda \equiv \rho/P$; so when $m(r)$ is not a constant, $\rho(r)/P(r)$ can be a more accurate replacement of $m(r)$. Then we can approximately rewrite the equation~(\ref{eqsta2}) as
% --------------------------------------------------------------------------------------
\begin{equation}
\frac{d\rho}{dr}=-\frac{dP}{5dr}(\lambda-4\frac{\rho}{P}),
\end{equation}
whose solution is
% --------------------------------------------------------------------------------------
\begin{equation}
\label{A3}
\rho = -\lambda P + \nu P^{4/5},
\end{equation}
where $\nu$ just is an integral constant. For the general value of $\nu$, if one wants to get a more exact solution, he can first denote (\ref{A3}) as $\rho^{(0)} = \rho^{(0)} (\lambda,\nu,P)$, then
% --------------------------------------------------------------------------------------
\begin{enumerate}
\item substitute $\rho^{(0)}$ into $m(r)$, which can be rewritten as $m(r)= m(\rho,P) = m(\lambda, \nu, P)$, then from equation~(\ref{eqsta2}) we can get a new density $\rho^{(1)} = \rho^{(1)}(\lambda, \nu, P)$;
\item repeat the step of (i), until the maximum of $|\rho^{(n+1)}(P)-\rho^{(n)}(P)|$  becomes smaller than $\epsilon$, which is the resolution we set.
\end{enumerate}
Notice that every time we iterate the function $\rho=\rho(\lambda,\nu,P)$, not the value of $\rho$ at a point. Then $\rho^{(n)}$ will be the solution of equation~(\ref{eqsta2}) with the resolution $\epsilon$.

% ------------------------------------------------------------------------------------
\begin{figure}
\centerline{\includegraphics[width=\columnwidth]{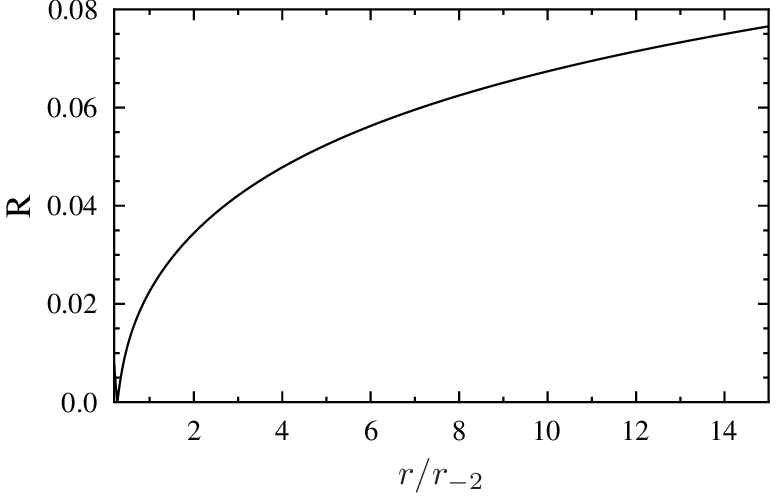}}
\caption{The relative error between $m$ and $\rho/P$ as a function of $r$.}
\label{resid}
\end{figure}

% ------------------------------------------------------------------------------------
However, equation~(\ref{A3}) has the main properties of equation~(\ref{eqsta2}). In fact, $m\sim\rho/P$ is really a good approximation except for large value of $\nu$, and hence the typical error of $m$ is small for small value of $\nu$. We define the relative error between $m$ and $\rho/P$ as $R = |(m - \rho/P)/m| = |1 - \rho/Pm|$, which is a function of $r$. We show $R$ vs. $r$ in Figure~\ref{resid}, in which all the data are taken from the results corresponding to equation~(\ref{A3}), i.e. equation~(\ref{sta:gal}), in Figure~\ref{dens1}. From Figure~\ref{dens1} we know that when $r/r_{-2}<0.2$, equation~(\ref{A3}) can be very well fitted by $\rho/P=-\lambda$, which almost ensures $R=0$, so we only focus on the external part. Notice that when $r > r_{-2}$, both $\rho$ and $P$ are very small, so even a small deviation from $\rho$ or $P$ may cause a high value of $R$. Figure~\ref{resid} indicates that with a chosen value of $\nu$, the approximation we adopted is very good. If we make more accurate approximation of $m(r)$, we can obtain more accurate solution of equation~(\ref{eqsta2}), but the shape of density profile will not be changed essentially.

% ---------------------------------------------------------------------------------
\end{document}